\documentclass[aps,prl,twocolumn]{revtex4}

\usepackage{color}
\usepackage{bm}
\usepackage{amsmath}
\usepackage{amssymb}
\usepackage{amsthm}
\usepackage{graphicx}
\usepackage{epsfig}
\usepackage{natbib}

\begin{document}

\title{Angle-resolved reflectance and surface plasmonics of the MAX phases}
\author{O. Kyriienko}
\affiliation{Science Institute, University of Iceland, Dunhagi 3,
IS-107, Reykjavik, Iceland}
\author{I. A. Shelykh}
\affiliation{Science Institute, University of Iceland, Dunhagi 3,
IS-107, Reykjavik, Iceland}
\affiliation{International Institute of Physics, Av. Odilon
Gomes de Lima, 1772, Capim Macio, 59078-400, Natal, Brazil}

\begin{abstract}
We investigate theoretically the optical response of bulk samples and thin films of the MAX phases materials, accounting for their large electrical anisotropy. We reveal the unusual behaviour of the reflection and transmittion spectra as a function of the incidence angle and predict the effect of the inverse total internal reflection. We also investigate the behaviour of the surface plasmon modes in bulk samples and thin films and analyse the difference between MAX materials and conventional metals.
\end{abstract}
\maketitle

\section{Introduction}
\begin{figure}
\includegraphics[width=1.0\linewidth]{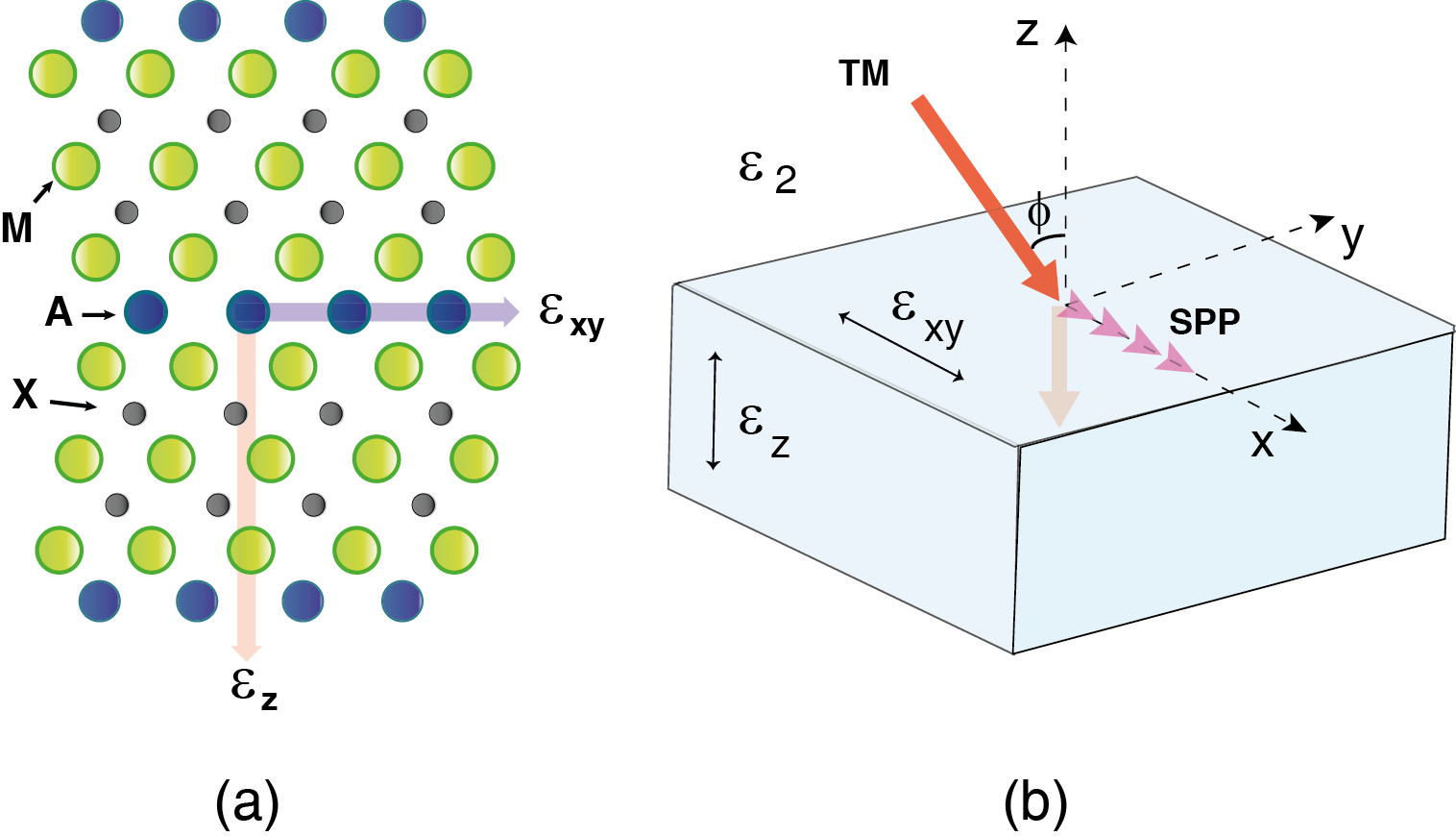}
\caption{(Color online) (a) Sketch of the MAX phase crystal structure in $xz$ plane. The conductivity in the system is provided by $A$ atoms and is greater in $x$ and $y$ directions. (b) Geometry of the system for semi-infinite dielectric-MAX phase interface for TM polarized beam. $\varepsilon_{2}$ denotes the isotropic dielectric constant of the media in contact with MAX phase, while $\varepsilon_{xy}$, $\varepsilon_{z}$ are components of dielectric tensor of MAX phase.}
\label{Fig1}
\end{figure}

The unique combination of metal and ceramics, the so-called MAX phases, attracts the growing interest of the material science community. It is a group of materials described by the general chemical formula $M_{n+1}AX_{n}$, where M denotes an early transition metal, A is a group IIIA or IVA element, X is either C and/or N and $n=1-3$\cite{Eklund,Toth}. The existence of over 50 such phases was revealed during the last decade. The MAX phases possess a number of unique physical properties making them suitable for a variety of practical implementations. Due to the anisotropic crystal structure (see Fig.\ref{Fig1}), they have extremely useful mechanical characteristics being stiff and possessing fully reversible deformation properties \cite{BarsoumAm,Barsoum}. They are good thermal and electrical conductors \cite{Eklund}. Moreover, since the MAX phases are layered structures, they demonstrate pronounced conductivity anisotropy: the conductivity along the crystalline layers can be several times bigger then across the layers \cite{Haddad}. Naturally, this should lead to peculiarities of the optical response of the bulk MAX materials and thin films.

Surprisingly, the optical properties of MAX phases still remain under-investigated from both experimental and theoretical points of view. From the theoretical side, there are \textit{ab-initio} calculations of the density of states in MAX phases and its effect on dielectric function of  Ti$_{3}$SiC$_{2}$ and Ti$_{3}$AlN$_{2}$ phases \cite{Haddad,Li}. From the experimental side, large conductivity anisotropy was detected using the optical methods in Ti$_{2}$AlC and Ti$_{2}$AlN phases showing different optical response regarding the crystal axis \cite{Haddad}. However, no detailed analysis of the angle resolved reflectivity spectra of MAX phases was performed up to now, to the best of our knowledge.

Considering good conductive properties of the MAX phases and their ideal mechanical properties, the analysis of the surface plasmonic modes in structure of their basis for creation of the plasmonic waveguides also becomes an actual task. Surface plasmon polaritons are confined electromagnetic states propagating across the metal-dielectric interface \cite{Pitarke}, which are of a great interest due to the possibility of building the ultra-dense and high-speed optoelectronic components \cite{Bozhevolniy}. The potential of the applications of MAX phases in this domain remains unexplored.

The current letter is aimed to bridge the gap in theoretical description of the optical properties of the bulk MAX phases and thin films on their basis. We analyze the reflectivity spectra, predicting the interesting effect of the inverse total internal reflection, and present a detailed consideration of the plasmonic modes.

\section{Reflectivity spectrum of bulk MAX structure}
The anisotropic nature of the MAX phases can be accounted for by introducing a dielectric permittivity tensor
\begin{equation}
\mathbf{\varepsilon} =
\left( \begin{array}{ccc}
\varepsilon_{xy} & 0 & 0  \\
0 & \varepsilon_{xy} & 0 \\
0 & 0 & \varepsilon_{z}
\end{array} \right)\,,
\end{equation}
where $\varepsilon_{xy}$ and $\varepsilon_{z}$ denote to the in-plane and perpendicular to plane dielectric constants respectively, which can be modelled using a standard Drude formula \cite{Pitarke}
\begin{equation}
\varepsilon_{xy,z}=1-\frac{\omega_{p}^{(xy,z)2}}{\omega(\omega+i\gamma)}\,,
\label{eps}
\end{equation}
where $\omega_{p}^{(xy,z)}$ denote to the plasma frequencies corresponding to in-plane and perpendicular to the plane directions and $\gamma$ describes the damping of the plasmons in the system. The difference between $\omega_{p}^{(xy)}$ and $\omega_{p}^{(z)}$ comes from the anisotropy of the conductivity, and it is possible to find a region of the frequencies for which $\varepsilon_{xy}$ is negative, but $\varepsilon_{z}$ is still positive. In particular, this situation was discovered for thin films and bulk samples of Ti$_{2}$AlC and Ti$_{2}$AlN studied in the Refs. \cite{Haddad,BarsoumMet}. E.g., the corresponding plasma frequencies for Ti$_{2}$AlC are $\hbar\omega_{p}^{(xy)}=20$ eV and $\hbar\omega_{p}^{(z)}=16$ eV with characteristic relaxation times in the range $\tau=2-0.2$ fs. In other MAX phases, however, the plasmonic frequencies can lie in the region of lower energies. From the point of view of the optical properties it means that in certain range of frequencies the system behaves as a metal in the $xy$ plain and as a dielectric in the $z$ direction. Let us consider the reflectivity as a function of the angle in this regime.

For the TE polarization of incident light the vector of the electric field lies in the $xy$ plane. Therefore, it can be considered as an ordinary wave in birefringent media with purely imaginary ordinary refractive index $n_{1}^{o} \rightarrow n_{xy} =i \sqrt{|\varepsilon_{xy}|}$. Naturally, this beam will be fully reflected from the semi-infinite media as from metallic surface if the damping can be neglected.
\begin{figure}
\includegraphics[width=1.0\linewidth]{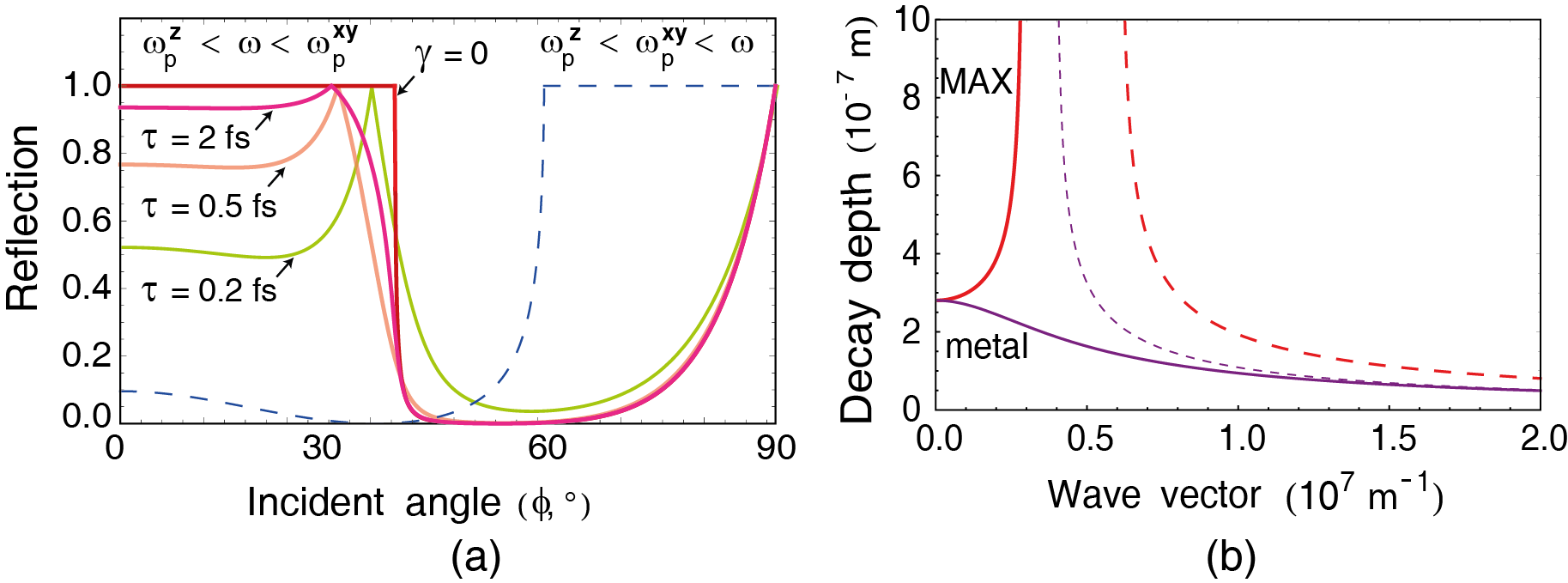}
\caption{(Color online) (a) Reflectivity of the MAX phase semi-infinite crystal for two different values of the frequency $\omega$. Red line corresponds to the case $\omega_{p}^{(z)}<\omega<\omega_{p}^{(xy)}$ (without damping) and demonstrates the inverse total internal reflection, while other solid lines show the influence of optical damping on reflectivity spectrum. Blue dashed line denotes to the case of fully dielectric behaviour and shows the effect of total internal reflection. (b) The attenuation length of the free electromagnetic wave in the MAX phase and metal for $\omega_{p}^{z}<\omega<\omega_{p}^{(xy)}$ (solid lines) and $\omega>\omega_{p}^{(xy,z)}$ (dashed lines).}
\label{Fig2}
\end{figure}

The case of TM polarized mode is much more interesting. In this case the electric field vector lies in the plane of incidence and thus the beam corresponds to extraordinary one in birefringent material, for which the effective refractive index depends on both $\varepsilon_{z}>0$ and $\varepsilon_{xy}<0$:
\begin{equation}
n_{1}^{e}=\sqrt{\varepsilon_{2}\sin^{2}\phi(1-\frac{\varepsilon_{xy}}{\varepsilon_{z}})+\varepsilon_{xy}}\,,
\label{ne}
\end{equation}
where the permittivity of second media $\varepsilon_2$ is taken to be unity in further discussion. Depending on the incidence angle, $n_{1}^{e}$ can be purely imaginary (small incidence angles $\phi$) or real (large incidence angles $\phi$). In the first case the beam is fully reflected from the interface, while in the second it can penetrate into the MAX phase. The reflection coefficients for TM polarized incident light as functions of the angle of incidence are plotted in the Fig. \ref{Fig2} (a). The solid lines show the reflection for frequency lying in the range $\omega_{p}^{(z)}<\omega<\omega_{p}^{(xy)}$. The red line denotes to the dissipationless case, while crimson, pink and green lines describe reflectance with accounting realistic optical damping rates $\gamma$ taken from the Ref. \cite{Haddad}. Dashed blue line shows the situation $\omega>\omega_{p}^{(xy,z)}$ where both dielectric constants $\varepsilon_{xy}$ and $\varepsilon_z$ are positive and response is fully dielectric. One sees that the behaviour is qualitatively different in these two cases. For fully dielectric response one clearly sees the effect of total internal reflection (reflectivity becomes unity above certain angle). For $\varepsilon_{xy}<0$ and $\varepsilon_{z}>0$ the opposite behaviour is manifested: full reflection for small incidence angles and transmission above critical angle $\theta_c=\arcsin[|\varepsilon_{xy}|\varepsilon_{z}/\varepsilon_{2}(\varepsilon_{z}+|\varepsilon_{xy}|)]$, the effect which can be considered as inverse total internal reflection. The account of the finite damping time leads to the partial absorption of the light by a MAX phase which leads to the deviation from perfect reflection at small angles, but the effect is still pronounced even in this case. 

Fig. \ref{Fig2}(b) describes the dependence of the penetration depth on the in-plane component of the wave vector of the incident light $q$. One clearly sees that in the transient region $\omega_{p}^{(z)}<\omega<\omega_{p}^{(xy)}$, the behaviour of the penetration depth in the MAX phase on $q$ is opposite to those in metal: it is an increasing function of $q$ for the first case and decreasing function in the second one.

\section{Surface plasmons}
Our next goal is to consider a surface plasmon polariton (SPP) appearing at the boundary between isotropic dielectric and anisotropic MAX phase media. It is a confined electromagnetic mode of TM polarization for which electric and magnetic fields decay exponentially in both directions with $\kappa_{i}$ being the inverse penetration depth to the media 1 (MAX phase) or 2 (isotropic dielectric). Introducing expressions for electric field and magnetic field vectors into the Maxwell equations and using standard boundary conditions for the components of the fields, one obtains straightforwardly the relation between $q$ and $\omega$ allowing to determine the dispersion of the surface plasmon at semi-infinite MAX phase--dielectric interface
\begin{equation}
q(\omega)=\frac{\omega}{c}\sqrt{\frac{\varepsilon_{2}\varepsilon_{z}(\omega)[\varepsilon_{2}-\varepsilon_{xy}(\omega)]}
{\varepsilon_{2}^{2}-\varepsilon_{xy}(\omega)\varepsilon_{z}(\omega)}},
\label{qMAX}
\end{equation}
where expressions (\ref{eps}) should be used for $\varepsilon_{xy,z}(\omega)$.
The decay constants are given by
\begin{eqnarray}\label{kappa1}
\kappa_1=\sqrt{\frac{\varepsilon_{xy}}{\varepsilon_{z}}(\omega)\left[q^{2}-\varepsilon_{z}\frac{\omega^{2}}{c^{2}}\right]},\\
\kappa_2=\sqrt{q^{2}-\varepsilon_{2}\frac{\omega^{2}}{c^{2}}}.
\label{kappa2}
\end{eqnarray}
\begin{figure}
\includegraphics[width=1.0\linewidth]{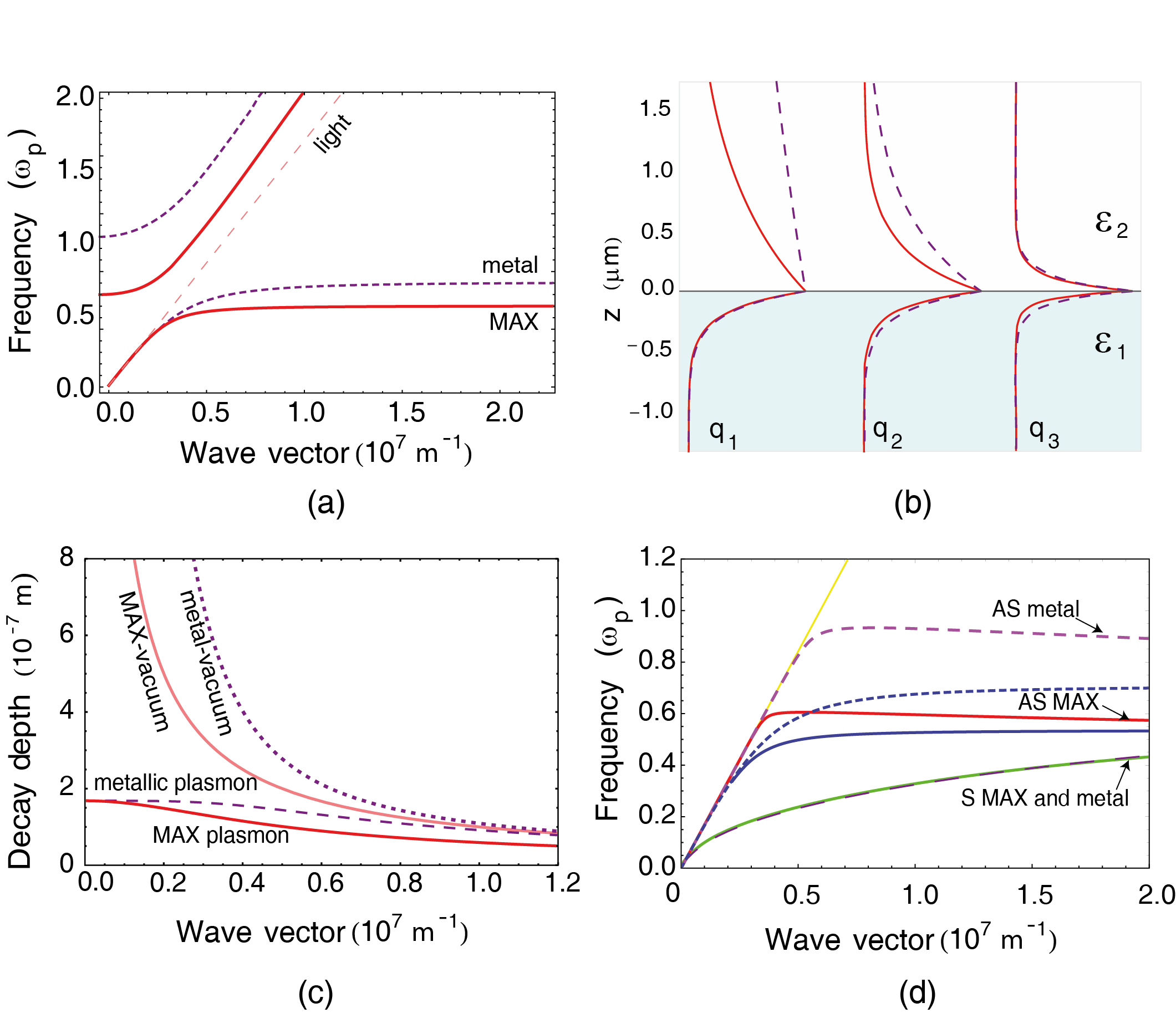}
\caption{(Color online) (a) Dispersion of SPP modes at the interface MAX phase-vacuum (red solid) and metal-vacuum (purple dashed). (b) The profiles of SPP modes at the interface MAX phase-vacuum (red solid) and metal-vacuum (purple dashed). (c) Decay length as a function of plasmon wave vector for MAX phase (red solid) and metal (purple dashed). (d) The dispersions of the symmetric and antisymmetric plasmonic modes in thin MAX phase film (solid) and metal (dashed) ($d=25$ nm). Blue lines correspond to the bulk surface plasmons.}
\label{Fig3}
\end{figure}
Fig. \ref{Fig3}(a) shows the dispersion of the surface plasmonic modes in semi-infinite MAX system compared to the case of the isotropic metal with $\omega_p=\omega_p^{(xy)}$. Fig. \ref{Fig3}(b) shows the spatial profile of the modes for three different values of $q$ ($q_{1}=10^{6} $m$^{-1}, q_{2}=3\times 10^{6} $m$^{-1}, q_{3}=10^{7} $m$^{-1}$) and Fig. \ref{Fig3}(c) shows the dependence of the penetration depths on $q$ in a plasmonic mode for the MAX phase and metallic media. One can note that for the MAX structure the surface plasmon decay length is smaller than for metal, which results into stronger concentration of electromagnetic field of a surface plasmon near the interface and leads to plasmon enhancement.

It is also possible to investigate surface plasmons in thin film samples. The task is actual since many MAX phases exist exactly in this configuration. The formal derivation requires writing boundary conditions for both sides of the film \cite{Economou}. Two confined modes corresponding to symmetric and antisymmetric distribution of the fields in the real space appear in this case. Their dispersions can be found from the following equations:
\begin{align}
\frac{\varepsilon_{xy}}{\kappa_{1}}+\frac{\varepsilon_{2}}{\kappa_{2}\tanh(\kappa_{1}d/2)}=0,\\
\frac{\varepsilon_{xy}}{\kappa_{1}}+\frac{\varepsilon_{2}}{\kappa_{2}\coth(\kappa_{1}d/2)}=0,
\label{cond-thin}
\end{align}
where $d$ is the thickness of the film and $\kappa_{1,2}$ are given by equations (\ref{kappa1})--(\ref{kappa2}).

The corresponding dispersions are plotted on the figure \ref{Fig3}(d) for the MAX phase and isotropic metallic film with $\omega_p=\omega_p^{(xy)}$. One sees that while dispersions of antisymmetric modes for MAX and metallic systems are clearly distinct, the symmetric modes coincide for both cases and show small differences only in high momenta region.

A great advantage of the MAX phases regarding usage in plasmonics devices as compare to the structures based on metamaterials \cite{Kaliteevski} is their unique mechanical and deformational properties. The specific kink deformation can allow the creation of flexible plasmonic wires of the desired geometry.  However, the investigation of this interesting configuration lies beyond the scope of present paper.

\section{Conclusions} In conclusion, we analysed the optical properties of MAX phases accounting for their conductivity anisotropy. For the case of different signs of dielectric permittivity tensor components $\varepsilon_{xy}$ and $\varepsilon_{z}$ we predicted the effect of the inverse total internal reflection. We also considered localized surface plasmon polariton modes in semi-infinite MAX structures and thin MAX phase films.

This work was supported by Rannis "Center of Excellence in Polaritonics". O.K. acknowledges a support from Eimskip foundation.


\begin{thebibliography}{99}

\bibitem{Eklund} P. Eklund, M. Beckers, U. Jansson, H. Hogberg, and L. Hultman, Thin Solid Films \textbf{518,}  1851-1878 (2010).

\bibitem{Toth} L. Toth, \textit{Transition Metal Carbides and Nitrides} (Academic, New York, 1971).

\bibitem{BarsoumAm} M. W. Barsoum and T. El-Raghy, Am. Sci. \textbf{89,} 337 (2001).

\bibitem{Barsoum} M. W. Barsoum, Progress in Solid State Chemistry \textbf{28,} 201-281 (2000).

\bibitem{Haddad} N. Haddad, E. Garcia-Caurel, L. Hultman, M. W. Barsoum, and
G. Hug, J. Appl. Phys. \textbf{104,} 023531 (2008).

\bibitem{Li} S. Li, R. Ahuja, M. W. Barsoum, P. Jena, and B. Johansson, Appl. Phys. Lett. \textbf{92,} 221907 (2008).

\bibitem{Pitarke} J. M. Pitarke, V. M. Silkin, E. V. Chulkov, and P. M. Echenique, Rep. Prog. Phys. \textbf{70,}  1-87 (2007).

\bibitem{Bozhevolniy} S. I. Bozhevolnyi, \textit{Plasmonic Nanoguides and Circuits} (Pan Stanford Publishing, Stanford, 2008).

\bibitem{BarsoumMet} M. W. Barsoum, M. Ali, and T. El-Raghy, Metall. Mater. Trans. A \textbf{31A}, 1857 (2000).

\bibitem{Economou} E. N. Economou, Phys. Rev. \textbf{182,} 539 (1969).

\bibitem{Kaliteevski} S. Brand, R. A. Abram, and M. A. Kaliteevski, Phys. Rev. B \textbf{75,} 035102 (2007).

\end{thebibliography}
\end{document}